\documentclass[aps,pre,reprint,superscriptaddress,twocolumn,longbibliography,floatfix]{revtex4-2}

\usepackage{empheq}
\usepackage{graphicx}
\usepackage{dcolumn}
\usepackage{bm}

\usepackage[utf8]{inputenc}
\usepackage[T1]{fontenc}
\usepackage{mathptmx}
\usepackage{etoolbox}
\usepackage{xcolor}
\usepackage{mathpazo}
\usepackage{bm}
\usepackage[unicode=true,
 bookmarks=true,bookmarksnumbered=false,bookmarksopen=false,
 breaklinks=false,pdfborder={0 0 0},pdfborderstyle={},backref=false,colorlinks=true]
 {hyperref}
\hypersetup{pdfborderstyle={},pdfborderstyle={},pdfborderstyle={},pdfborderstyle={},pdfborderstyle={},pdfborderstyle={},linkcolor=blue,citecolor=blue}
\usepackage{breakurl}

\usepackage{natbib}
\usepackage{amssymb,amsmath}
\usepackage{comment}

\usepackage{float}
\newcommand{\tx}{\widetilde{x}}
\newcommand{\tg}{\widetilde{g}}
\newcommand{\tG}{\widetilde{G}}
\newcommand{\tD}{\widetilde{D}}
\newcommand{\tge}{\widetilde{g}^{\text{env}}}
\newcommand{\phib}{\bar{\phi}}

\newcommand{\bp}{\beta p}

\newcommand\beq{\begin{equation}}
	\newcommand\eeq{\end{equation}}
\newcommand\beqa{\begin{eqnarray}}
	\newcommand\eeqa{\end{eqnarray}}
\newcommand{\nn}{\nonumber\\}
\def\bal#1\eal{\begin{align}#1\end{align}}

\makeatother
\begin{document}

\title[Algebraic to exponential decay of spatial correlations]{Algebraic to exponential decay of spatial correlations in one-dimensional
 and confined hard-core fluids: A Laplace-pole analysis}
\author{Ana M. Montero}
\email{anamontero@unex.es}
\affiliation{Departamento de F\'isica, Universidad de Extremadura, E-06006 Badajoz, Spain}

\author{Andr\'es Santos}
\affiliation{Departamento de F\'isica, Universidad de Extremadura, E-06006 Badajoz, Spain}
\affiliation{Instituto de Computaci\'on Cient\'ifica Avanzada (ICCAEx), Universidad de Extremadura, E-06006 Badajoz, Spain}

\date{\today}

\begin{abstract}
We derive the asymptotic behavior of the radial distribution function $g(x)$ for one-dimensional (1D) hard-rod systems and related quasi-one-dimensional geometries at high packing fractions using Laplace transform techniques and pole analysis. By identifying the poles and residues of the Laplace transform in the limit of small void fraction, we obtain compact representations of $g(x)$ in terms of the Jacobi elliptic theta function $\theta_3$. This formulation naturally captures the two regimes governing the oscillatory decay toward unity: an intermediate algebraic decay and a long-distance exponential decay, consistent with previous results for the Tonks gas. Our approach provides a unified framework that (i) expresses $g(x)$ in a single well-tabulated special function, (ii) links spatial correlations directly to the pole structure in complex Laplace space, offering clear physical insight into decay rates and oscillation frequencies, and (iii) generalizes straightforwardly to 1D binary mixtures and confined hard-disk systems, where direct Gaussian decompositions are cumbersome. The equivalence between the theta-function representation and the Gaussian superposition of Bouzar and Messina [Phys. Rev. E \textbf{112}, L042105 (2025)] is established via the Poisson summation formula, highlighting the versatility and conceptual advantages of the Laplace-pole framework.
\end{abstract}

\maketitle

\section{Introduction}

In a recent paper \cite{BM25}, Bouzar and Messina revisited the exact radial distribution function (RDF) $g(x)$ of the one-dimensional (1D) hard-rod fluid (Tonks gas) \cite{T36,S16}.
Although the Tonks gas is a well-known and extensively studied system, their work revealed noteworthy features, underscoring that, despite its apparent simplicity, it remains relevant both theoretically and as an effective model for single-file confined hard disks and hard spheres \cite{MS24b,MS25}.

In particular, Bouzar and Messina showed that, when the packing fraction $\phi$ is sufficiently close to its close-packing value $\phi_{\text{cp}}=1$, the oscillatory decay of $g(x)$ exhibits two distinct regimes: (i) for $(x-1)\phi\lesssim 0.077/\phib^2$, the decay is algebraic with an envelope $g^{\text{env}}(x)\sim(x-1)^{-1/2}$, whereas (ii) for $(x-1)\phi\gtrsim 0.077/\phib^2$, the decay is exponential with an envelope $g^{\text{env}}(x)-1\sim e^{-\kappa x}$. Here the rod length is taken as the unit of length, $\phib\equiv 1-\phi$ denotes the void fraction, and $\kappa=2\pi^2\phi\phib^2$ is the damping coefficient. The algebraic regime (i) signals the emergence of crystalline order, which is eventually frustrated by the exponential regime (ii), except in the limiting case $\phi\to\phi_{\text{cp}}$.

In their analysis, Bouzar and Messina introduced the scaled variable
\beq
\label{1}
\tx=(x-1)\phi
\eeq
and showed that, for sufficiently  small $\phib$,  $g(x)$ can be accurately approximated  by
\beq
\tg(\tx)\equiv g(x)\approx \phib^{-1}\sum_{n=-\infty}^\infty \frac{e^{-{(\tx-n)^2}/{2\phib^{2}|n|}}}{\sqrt{2\pi|n|}},
\label{gauss}
\eeq
which has local maxima at $\tx=\text{integer}$, with an envelope
\beq
\label{envelope}
\tge(\tx)=\theta_3\left(0,e^{-2\pi^2\phib^2\tx}\right),
\eeq
where
\bal
\label{theta3}
\theta_3(z,q)=&\sum_{n=-\infty}^\infty q^{n^2}e^{2 \imath nz}\nn=&
1+2\sum_{n=1}^\infty q^{n^2}\cos(2  nz),\quad |q|<1,
\eal
is the Jacobi elliptic theta function \cite[\href{https://dlmf.nist.gov/20.2.E3}{(20.2.3)}]{DLMF}. Taking into account its mathematical properties
\beq
\label{theta0}
\theta_3(0,q)\simeq
\begin{cases}
\displaystyle{\sqrt{\frac{\pi}{\ln q^{-1}}}}, & 0.217\lesssim q<1\\
 1+2q,&0\leq q \lesssim 0.217,
\end{cases}
\eeq
the two  regimes (i) and (ii) follow directly as
\beq
\label{alg}
\tge(\tx) \simeq
\begin{cases}
\displaystyle{\frac{1}{\sqrt{2\pi\phib^2\tx}}},& \displaystyle{\tx\lesssim \frac{0.077}{\phib^2}}\\
\displaystyle{1+2e^{-2\pi^2\phib^2\tx}},&\displaystyle{\tx\gtrsim \frac{0.077}{\phib^2}}.
\end{cases}
\eeq
The intermediate algebraic decay in Eq.~\eqref{alg} had already been obtained by Hu and Charbonneau~\cite{HC21} through a simple physical argument.

{\renewcommand{\arraystretch}{2}
	\begin{table*}
		\caption{Summary and comparison of the main results obtained in this work for the Tonks gas, the symmetric nonadditive binary mixture, and confined hard disks. The table compiles the scaling definitions, pole locations, theta-function forms of the RDFs and their envelopes, and the associated algebraic and exponential asymptotic behaviors in the high-packing-fraction limit.\label{tab1}}
		\begin{ruledtabular}
			\begin{tabular}{cccc}
				Property&Tonks gas&Binary mixture&Confined hard disks\\
				\hline
				$\phi$&$\rho$&$\rho a$&$\rho a, \quad a=\sqrt{1-\epsilon^2}$\\
				$\tx$&$(x-1)\phi$&$\displaystyle{\left(\frac{x}{a}-1\right)\phi}$&$\displaystyle{\left(\frac{x}{a}-1\right)\phi}$\\
				$\zeta_n$&$2n^2\pi^2\phib^2$&$\displaystyle{\frac{n^2\pi^2\phib^2}{2}}$&$\displaystyle{\frac{n^2\pi^2\phib^2}{4}}$\\
				$\omega_n$&$2n\pi$&$n\pi$&$n\pi$\\
				$\tg(\tx)$&$\displaystyle{\theta_3\left(\pi\tx,e^{-2\pi^2\phib^2\tx}\right)}$&$\displaystyle{\theta_3\left(\pi\tx,e^{-2\pi^2\phib^2\tx}\right)}$&$\displaystyle{\theta_3\left(\pi\tx,e^{-\pi^2\phib^2\tx}\right)}$\\
				$\tge(\tx)$&$\displaystyle{\theta_3\left(0,e^{-2\pi^2\phib^2\tx}\right)}$&$\displaystyle{\theta_3\left(0,e^{-2\pi^2\phib^2\tx}\right)}$&$\displaystyle{\theta_3\left(0,e^{-\pi^2\phib^2\tx}\right)}$\\
				Algebraic decay of $\tge(\tx)$&$\displaystyle{\frac{1}{\sqrt{2\pi\phib^2\tx}}},\quad\displaystyle{\tx\lesssim \frac{0.077}{\phib^2}}$&$\displaystyle{\frac{1}{\sqrt{2\pi\phib^2\tx}}},\quad\displaystyle{\tx\lesssim \frac{0.077}{\phib^2}}$&$\displaystyle{\frac{1}{\sqrt{\pi\phib^2\tx}}},\quad\displaystyle{\tx\lesssim \frac{0.15}{\phib^2}}$\\
				Exponential decay of $\tge(\tx)-1$&$\displaystyle{2e^{-2\pi^2\phib^2\tx}}, \quad\displaystyle{\tx\gtrsim \frac{0.077}{\phib^2}}$&$\displaystyle{2e^{-2\pi^2\phib^2\tx}}, \quad\displaystyle{\tx\gtrsim \frac{0.077}{\phib^2}}$&$\displaystyle{2e^{-\pi^2\phib^2\tx}}, \quad\displaystyle{\tx\gtrsim \frac{0.15}{\phib^2}}$\\
				$\tg_{11}(\tx),\quad \tg_{++}(\tx)$&---&$\displaystyle{\theta_3\left(\frac{\pi(\tx-1)}{2},e^{-\pi^2\phib^2\tx/2}\right)}$&$\displaystyle{\theta_3\left(\frac{\pi(\tx-1)}{2},e^{-\pi^2\phib^2\tx/4}\right)}$\\
				$\tg_{12}(\tx),\quad \tg_{+-}(\tx)$&---&$\displaystyle{\theta_3\left(\frac{\pi\tx}{2},e^{-\pi^2\phib^2\tx/2}\right)}$&$\displaystyle{\theta_3\left(\frac{\pi\tx}{2},e^{-\pi^2\phib^2\tx/4}\right)}$\\
				$\tge_{ij}(\tx)$&---&$\displaystyle{\theta_3\left(0,e^{-\pi^2\phib^2\tx/2}\right)}$&$\displaystyle{\theta_3\left(0,e^{-\pi^2\phib^2\tx/4}\right)}$\\
				Algebraic decay of $\tge_{ij}(\tx)$&---&$\displaystyle{\frac{2}{\sqrt{2\pi\phib^2\tx}}},\quad\displaystyle{\tx\lesssim \frac{0.31}{\phib^2}}$&$\displaystyle{\frac{2}{\sqrt{\pi\phib^2\tx}}},\quad\displaystyle{\tx\lesssim \frac{0.62}{\phib^2}}$\\
				Exponential decay of $\tge_{ij}(\tx)-1$&---&$\displaystyle{2e^{-\pi^2\phib^2\tx/2}}, \quad\displaystyle{\tx\gtrsim \frac{0.31}{\phib^2}}$&$\displaystyle{2e^{-\pi^2\phib^2\tx/4}}, \quad\displaystyle{\tx\gtrsim \frac{0.62}{\phib^2}}$\\
			\end{tabular}
		\end{ruledtabular}
	\end{table*}

While the Gaussian representation \eqref{gauss} offers valuable physical insight by decomposing the RDF into localized contributions, alternative mathematical formulations can provide complementary advantages. In statistical mechanics, different representations of the same quantity often highlight distinct aspects of the underlying physics and suggest different routes for generalization or approximation. In particular, Fourier and Laplace representations have proven especially useful in the analysis of correlation functions, as they naturally expose the characteristic frequencies and decay rates associated with spatial oscillations.

Despite these advances, important aspects of the correlation structure near close packing remain unexplored. In this paper, we extend the  results of Refs.~\cite{BM25,HC21} using a Laplace transform combined with a pole analysis. This approach leads to a compact representation in terms of the Jacobi theta function $\theta_3$, which is practically equivalent to Eq.~\eqref{gauss} but offers several practical and conceptual advantages: (a) it involves a single, well-tabulated special function rather than an infinite sum, (b) it directly links the spatial oscillation structure to poles in the complex frequency plane, clarifying the underlying decay mechanisms, and (c) it applies straightforwardly to related systems, including 1D mixtures and quasi-one-dimensional (q1D) geometries, where the direct Gaussian decomposition becomes algebraically cumbersome.
Moreover, unlike previous approaches, the present formulation yields explicit expressions not only for the envelope but also for the full oscillatory structure of the correlation functions, including partial RDFs in mixtures and confined geometries.

In addition to reproducing the Tonks-gas asymptotics, the present work provides a unified and explicit description of the full correlation functions in 1D binary mixtures and q1D confined hard-disk systems, allowing a direct comparison of decay rates, oscillation frequencies, and crossover scales across these models (see Table~\ref{tab1}).

The common origin of this behavior in the three systems considered here is the emergence of an effectively one-dimensional ordered arrangement near close packing. In the Tonks gas, particles naturally form a single-file sequence. In the symmetric nonadditive binary mixture, negative nonadditivity favors an alternating configuration $1$-$2$-$1$-$2$-$\cdots$, which makes the system behave effectively as a monodisperse lattice at high density. Similarly, for confined hard disks, the disks organize into a zigzag structure in which successive particles occupy opposite transverse positions, again producing an alternating quasi-one-dimensional arrangement. A closely related mechanism also appears in systems of freely rotating hard dumbbells under strong confinement, where the orientational ordering generates the same type of intermediate algebraic decay before the final exponential damping \cite{MGVS26}. Nevertheless, the equivalence with the Tonks gas is not exact: the effective nearest-neighbor spacing, the close-packing equation of state, and the pole damping coefficients differ from one system to another, leading to the quantitative differences summarized in Table~\ref{tab1}.

The paper is organized as follows. Section~\ref{sec2} presents the Laplace-transform and pole analysis for the monocomponent hard-rod system (Tonks gas). The same strategy is then applied to the more general case of a symmetric nonadditive binary mixture. In Sec.~\ref{sec4}, we move beyond strictly 1D systems and analyze the RDFs of hard disks under single-file confinement. Finally, Sec.~\ref{sec5} summarizes our conclusions.

\section{Monocomponent system}
\label{sec2}

We introduce the Laplace representation
\beq
G(s)=\int_0^\infty dx\, e^{-s x} g(x)
\eeq
of $g(x)$. For the Tonks gas, this transform is \cite{S16}
\beq
G(s)=\phi^{-1}\left[e^s\left(1+\frac{\phib }{\phi}s\right)-1\right]^{-1}.
\eeq
The Laplace transform of $\tg(\tx)$ is related to $G(s)$ through
\beq
\label{7}
\tG(s)=\phi e^{\phi s}G\left(\phi s\right)
=\left(1+\phib s-e^{-\phi s}\right)^{-1}.
\eeq

We now denote by $\{s_n=-\zeta_n\pm \imath\omega_n, \, n=1,2,\ldots\}$ the nonzero poles of $\tG(s)$.
Their real and imaginary parts satisfy
\beq
\label{Real}
1-\phib\zeta_n=e^{\phi \zeta_n}\cos\left(\phi\omega_n\right),\quad -\phib\omega_n=e^{\phi \zeta_n}\sin\left(\phi\omega_n\right),
\eeq
and the corresponding residues are
\bal
A_n=&\lim_{s\to s_n}(s-s_n)\tG(s)=\left(1+\phi  \phib s_n\right)^{-1}\nn
=&\frac{1-\phi\phib(\zeta_n\pm\imath\omega_n)}{\left(1-\phi\phib\zeta_n\right)^2+\left(\phi\phib\omega_n\right)^2}.
\eal
Thus $A_n=|A_n|e^{\pm\imath\delta_n}$, with
\begin{subequations}
\beq
|A_n|=\left[\left(1-\phi\phib\zeta_n\right)^2+\left(\phi\phib\omega_n\right)^2\right]^{-1/2},
\eeq
\beq
\delta_n=\mp\tan^{-1}\frac{\phi\phib\omega_n}{1-\phi\phib\zeta_n}.
\eeq
\end{subequations}
In terms of these poles and the residues, $\tg(\tx)$ admits the exact representation
\bal
\label{9}
\tg(\tx)=&1+\sum_{n=1}^\infty A_n e^{s_n \tx}\nn
=&1+2\sum_{n=1}^\infty|A_n|e^{-\zeta_n \tx}\cos(\omega_n \tx+\delta_n).
\eal

\begin{figure}
      \includegraphics[width=0.8\columnwidth]{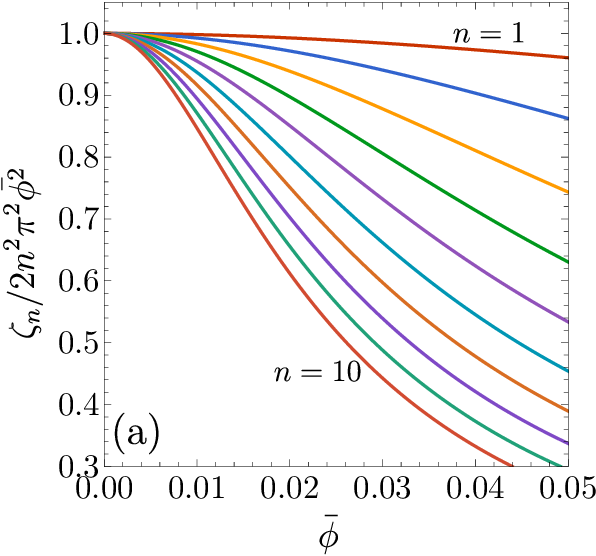}\\ \includegraphics[width=0.8\columnwidth]{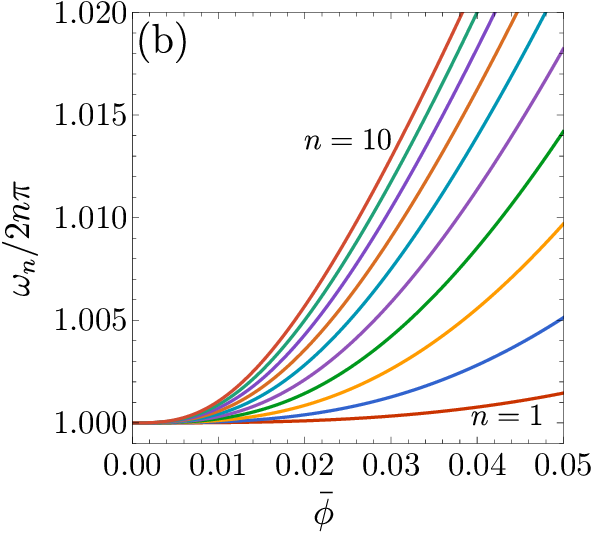}
           \caption{Monocomponent system. (a) $\zeta_n/2n^2\pi^2\phib^2$ and (b) $\omega_n/2n\pi$ as functions of $\phib$ for $n=1,2,\ldots,10$ [from top to bottom in panel (a) and from bottom to top in panel (b)].
\label{fig1}}
\end{figure}

In general, the coupled equations \eqref{Real} must be solved numerically. However, in the limit of small void fraction ($\phib\ll 1$), their solutions reduce to
\begin{subequations}
\beq
\label{11a}
\zeta_n=  2n^2\pi^2\phib^2+\mathcal{O}(\phib^{4}),\quad \omega_n= 2n\pi+\mathcal{O}(\phib^{3}),
\eeq
\beq
|A_n|=1+\mathcal{O}(\phib^{2}),\quad
\delta_n=\mathcal{O}(\phib).
\eeq
\end{subequations}
Inserting these expressions into Eq.~\eqref{9} gives
\bal
\tg(\tx)\approx &1+2\sum_{n=1}^\infty e^{-2n^2\pi^2\phib^2\tx}
\cos (2n\pi \tx)\nn
=&\theta_3\left(\pi \tx,e^{-2\pi^2\phib^2\tx}\right),\quad \phib\ll 1,
\label{Myg}
\eal
where the second step follows from Eq.~\eqref{theta3}.

Figure \ref{fig1} shows the ratios $\zeta_n/2n^2\pi^2\phib^2$ and $\omega_n/2n\pi$ for $0\leq\phib\leq 0.05$ and $n=1,2,\ldots,10$. As expected, the accuracy of the asymptotic expressions \eqref{11a} deteriorates as $n$ and/or $\phib$ increase.
Nevertheless, Eq.~\eqref{Myg} remains remarkably accurate, as illustrated in Fig.~\ref{fig2}.

\begin{figure}
      \includegraphics[width=0.8\columnwidth]{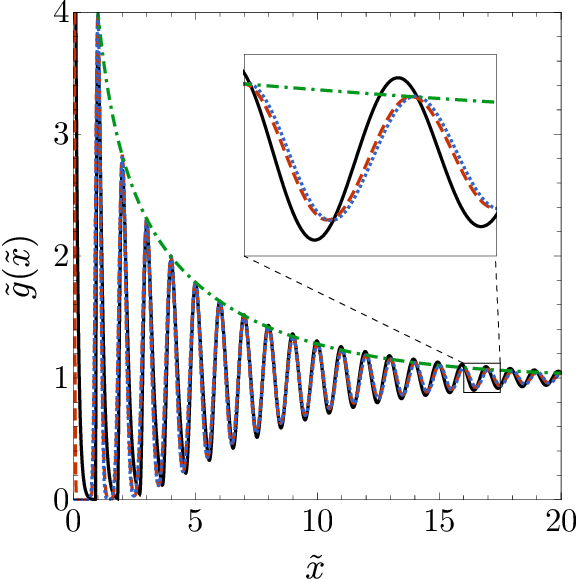}
            \caption{Monocomponent system. Plot of $\tg(\tx)$ for $\phi=0.9$. The solid, dotted, and dashed lines correspond to the exact function, the Gaussian approximation [Eq.~\eqref{gauss}], and our approximation [Eq.~\eqref{Myg}], respectively. The envelope [Eq.~\eqref{envelope}] is also shown (dashed-dotted line). The dashed and dotted lines are nearly indistinguishable.
\label{fig2}}
\end{figure}

It can be verified that the local maxima of $\theta_3(z, e^{-az})$ occur at $z_m\approx m\pi-\mathcal{O}(a)$, with $m=0,1,2,\ldots$. Consequently, for small $a$, the envelope of $\theta_3(z, e^{-az})$ is well approximated by $\theta_3(0, e^{-az})$.
Thus, the representation in Eq.~\eqref{Myg} leads to the same conclusions as those obtained by Bouzar and Messina from Eq.~\eqref{gauss}.
Indeed, using the Poisson summation formula $\sum_{n=-\infty}^\infty f(n)=\sum_{n=-\infty}^\infty\int_{-\infty}^\infty dx\, f(x)e^{-2\pi\imath nx}$ \cite[\href{https://dlmf.nist.gov/1.8.E14}{(1.8.14)}]{DLMF}, one finds
\beq
\label{Poisson}
\theta_3\left(\pi \tx,e^{-2\pi^2\phib^2\tx}\right)
=\frac{\phib^{-1}}{\sqrt{2\pi \tx}}\sum_{n=-\infty}^\infty
e^{-{(\tx-n)^{2}}/{2\phib^2 \tx}}.
\eeq
Approximating $e^{-{(\tx-n)^{2}}/{2\phib^2 \tx}}/\sqrt{2\pi \tx} \approx e^{-{(\tx-n)^{2}}/{2\phib^2 |n|}}/\sqrt{2\pi |n|}$ then recovers Eq.~\eqref{gauss} from Eqs.~\eqref{Myg} and \eqref{Poisson}.

\section{Symmetric nonadditive binary mixtures}
\label{sec3}
We now consider an equimolar binary hard-rod mixture with diameters $\sigma_{11}=\sigma_{22}=1$ and $\sigma_{12}=a<1$ (negative nonadditivity). It can be shown \cite{S16} that the equation of state is
\beq
\label{14}
\frac{1}{\rho}=\frac{1}{\bp}+\frac{a+e^{-(1-a)\bp}}{1+e^{-(1-a)\bp}},
\eeq
where $\rho$ is the number density, $\beta=1/k_BT$ is the inverse temperature, and $p$ is the pressure.
In general, $\bp$ cannot be expressed explicitly as a function of $\rho$.
However, in the high-pressure regime ($\bp\gg 1$),  the equation of state gives  $\bp a\approx{\phi}/{\phib}$ with $\phi=\rho a$ and $\phib=1-\phi$.

By analogy with Eq.~\eqref{1}, we define
\beq
\label{1new}
\tx=\left(\frac{x}{a}-1\right)\phi
\eeq
and $\tg_{ij}(\tx)\equiv g_{ij}(x)$. The Laplace transform of  $\tg_{ij}(\tx)$ is $\tG_{ij}(s)=(\phi/a)e^{\phi s}G_{ij}(\phi s/a)$, where $G_{ij}(s)$ is the Laplace transform of $g_{ij}(x)$ (see Appendix \ref{appA}). Consequently,
\begin{subequations}
\beq
\tG_{11}(s)\approx 2\frac{1-\tD(s)}{\tD(s)}e^{\phi s},\quad \tG_{12}(s)\approx\frac{2}{\left(1+\phib s\right)\tD(s)},
\eeq
\beq
\label{21c}
\tD(s)\approx\frac{1+{\phib s}-e^{- \phi s}}{1+\phib s}\frac{1+{\phib s}+e^{- \phi s}}{1+\phib s}.
\eeq
\end{subequations}
The poles $\{s_n=-\zeta_n\pm\imath\omega_n\}$ of $\tG_{ij}(s)$ are the roots of $\tD(s)=0$, i.e.,
\beq
\label{22}
1+{\phib s_n}=\pm e^{- \phi s_n}.
\eeq
Comparison with Eq.~\eqref{7} shows that the roots with  the  $+$ sign coincide with those of the monocomponent system,
\beq
\label{23}
\zeta_n\approx  \frac{n^2\pi^2\phib^2}{2},\quad \omega_n\approx n\pi,
\eeq
with $n=2,4,\ldots=\text{even}$. The roots corresponding to the  $-$ sign are also given by Eq.~\eqref{23}, but now with $n=1,3,\ldots=\text{odd}$.

\begin{figure}
      \includegraphics[width=0.8\columnwidth]{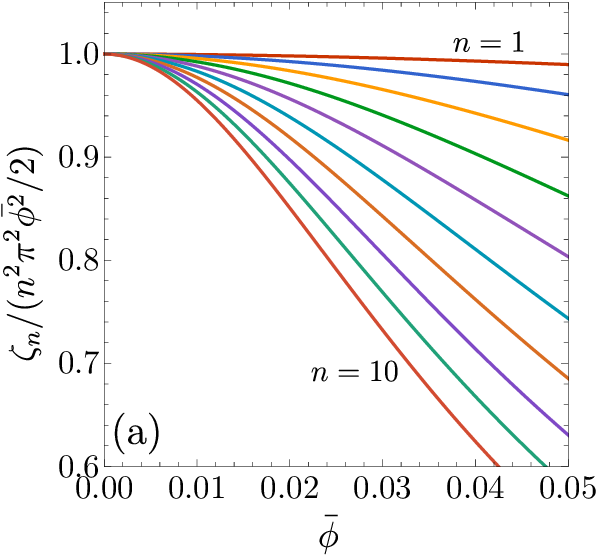}\\ \includegraphics[width=0.8\columnwidth]{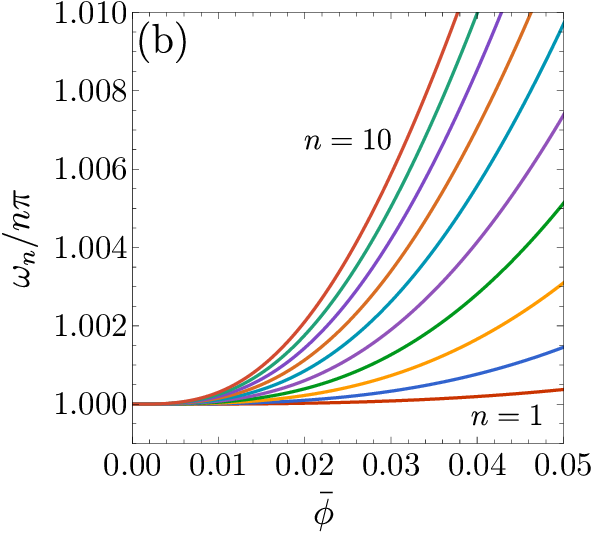}
            \caption{Symmetric binary mixture with $a=\frac{1}{2}$. (a) $\zeta_n/(n^2\pi^2\phib^2/2)$ and (b) $\omega_n/n\pi$ as functions of $\phib$ for $n=1,2,\ldots,10$ [from top to bottom in panel (a) and from bottom to top in panel (b)].
  \label{fig3}}
\end{figure}

\begin{figure}
      \includegraphics[width=0.8\columnwidth]{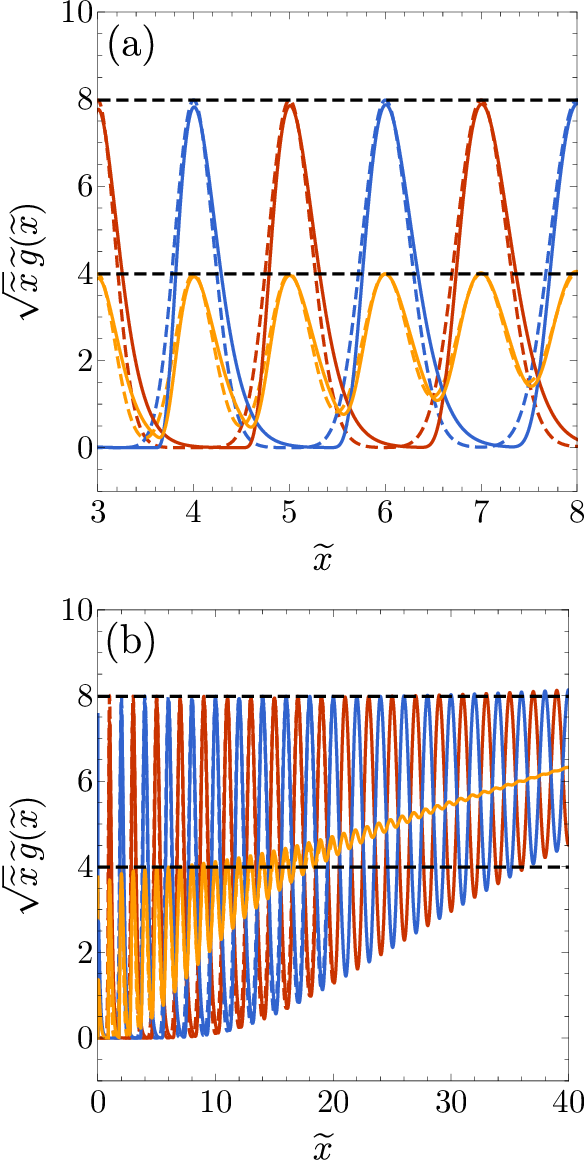}
            \caption{Symmetric binary mixture with $a=\frac{1}{2}$. Plot of $\sqrt{\tx}\tg_{11}(\tx)$ (maxima at $\tx\approx 1,3,5,\ldots$), $\sqrt{\tx}\tg_{12}(\tx)$ (maxima at $\tx\approx 0,2,4,\ldots$), and $\sqrt{\tx}\tg(\tx)$ (maxima at $\tx\approx 0,1,2,\ldots$) for $\phi=0.9$. Solid lines correspond to the exact functions, and dashed lines to our approximations [Eqs.~\eqref{25}]. The horizontal dashed lines indicate $2/\sqrt{2\pi\phib^2}\simeq 7.98$ and $1/\sqrt{2\pi\phib^2}\simeq 3.99$. (a) $3\leq\tx\leq 8$. (b) $0\leq\tx\leq 40$.
  \label{fig4}}
\end{figure}

The ratios $\zeta_n/(n^2\pi^2\phib^2/2$) and $\omega_n/n\pi$ for $n=1,2,\ldots,10$ are shown in Fig.~\ref{fig3} for a binary mixture with $a=\frac{1}{2}$. The behavior is qualitatively similar to that of the monocomponent fluid (cf. Fig.~\ref{fig1}), although the deviations from unity are smaller in the mixture.

To compute the residues, note that
\beq
\tD'(s_n)=2\frac{1+\phi\phib s_n}{1+\phib s_n}\approx 2.
\eeq
Hence,
\begin{subequations}
\beq
\label{25a}
\lim_{s\to s_n}(s-s_n)\tG_{11}(s)=\left(-1\right)^n,
\eeq
\beq
\lim_{s\to s_n}(s-s_n)\tG_{12}(s)=1.
\eeq
\end{subequations}
In Eq.~\eqref{25a} we have used that $e^{-\phi s_n}=(-1)^n(1+\phib s_n)\approx (-1)^n$.

Finally, applying the residue theorem, we obtain
\begin{subequations}
\label{25}
\bal
\tg_{11}(\tx)\approx& 1+2\sum_{n=1}^\infty \left(-1\right)^n e^{-n^2\pi^2\phib^2\tx/2}\cos\left( n\pi\tx\right)
\nn
=&\theta_3\left(\frac{\pi(\tx-1)}{2},e^{-\pi^2\phib^2\tx/2}\right),
\eal
\bal
\tg_{12}(\tx)\approx& 1+2\sum_{n=1}^\infty  e^{-n^2\pi^2\phib^2\tx/2}\cos\left( n\pi\tx\right)\nn
=&\theta_3\left(\frac{\pi\tx}{2},e^{-\pi^2\phib^2\tx/2}\right),
\eal
\bal
\tg(\tx)=& \frac{1}{2}\left[\tg_{11}(\tx)+\tg_{12}(\tx)\right]\nn
\approx&1+2\sum_{n=1}^\infty  e^{-2n^2\pi^2\phib^2\tx}\cos\left( 2n\pi\tx\right)\nn
=&\theta_3\left(\pi\tx,e^{-2\pi^2\phib^2\tx}\right).
\eal
\end{subequations}
The corresponding envelopes are
\begin{subequations}
\beq
\tge_{11}(\tx)
=\tge_{12}(\tx)=\theta_3\left(0,e^{-\pi^2\phib^2\tx/2}\right),
\eeq
\beq
\tge(\tx)=\theta_3\left(0,e^{-2\pi^2\phib^2\tx}\right).
\eeq
\end{subequations}

It is worth noting that only the even poles of $\tg_{ij}(\tx)$ contribute to the total RDF $\tg(\tx)$. Consequently, while $\tg_{11}(\tx)$ and $\tg_{12}(\tx)$ have local maxima at $\tx\approx 1,3,5,\ldots$ and $\tx\approx 0,2,4,\ldots$, respectively, $\tg(\tx)$ exhibits local maxima at $\tx\approx 0,1,2,\ldots$. This reflects the near-close-packing particle arrangement of alternating types: $1$-$2$-$1$-$2$-$\cdots$.
Equation~\eqref{theta0} then implies
\begin{subequations}
\beq
\label{alg11}
\tge_{ij}(\tx) \simeq
\begin{cases}
\displaystyle{\frac{2}{\sqrt{2\pi\phib^2\tx}}},& \displaystyle{\tx\lesssim \frac{0.31}{\phib^2}}\\
\displaystyle{1+2e^{-\pi^2\phib^2\tx/2}},&\displaystyle{\tx\gtrsim \frac{0.31}{\phib^2}},
\end{cases}
\eeq
\beq
\label{algbin}
\tge(\tx) \simeq
\begin{cases}
 \displaystyle{\frac{1}{\sqrt{2\pi\phib^2\tx}}},& \displaystyle{\tx\lesssim \frac{0.077}{\phib^2}}\\
\displaystyle{ 1+2e^{-2\pi^2\phib^2\tx}},&\displaystyle{\tx\gtrsim \frac{0.077}{\phib^2}}.
\end{cases}
\eeq
\end{subequations}

The overall behavior of the total RDF $\tge(\tx)$ closely resembles that of the monocomponent system. However, the algebraic decay of the partial functions $\tge_{ij}(\tx)$ extends to distances four times larger than that of $\tge(\tx)$.

The features described above are clearly illustrated in Fig.~\ref{fig4}. In particular, the algebraic decay of the total RDF, $\tg(\tx)\sim 1/\sqrt{\tx}$, breaks down for $\tx\gtrsim 10$, whereas the algebraic decay of the partial functions, $\tg_{ij}(\tx)$, persists up to $\tx\gtrsim 40$.

\section{Extension to confined hard disks}
\label{sec4}

We now consider  a q1D system of hard disks of unit diameter confined between two parallel lines. The disks are free to move along the longitudinal axis $x$, while their transverse positions are restricted to $-\frac{\epsilon}{2} \leq y \leq \frac{\epsilon}{2}$, where $1+\epsilon$ is the wall separation. For $\epsilon \leq \frac{\sqrt{3}}{2}$, each disk interacts only with its immediate left and right neighbors \cite{KP93}.

It can be shown that this q1D system of identical disks can be mapped onto a 1D polydisperse system of nonadditive hard rods \cite{MS23,MS23b,MS24,ThesisAna}, which is considerably more complex than the binary mixture analyzed in Sec.~\ref{sec3}.

The longitudinal distance of two disks in contact, with transverse coordinates $y$ and $y'$, is $\sigma(y,y')=\sqrt{1-(y-y')^2}$, which ranges from $\sigma(y,y)=1$ to $\sigma(\pm\frac{\epsilon}{2},\mp\frac{\epsilon}{2})=\sqrt{1-\epsilon^2}\equiv a$. The RDF, $g(x;y,y')$, thus depends on the longitudinal separation $x$ and the transverse positions $y$ and $y'$ of both particles. The total RDF is obtained by averaging over all transverse positions:
\beq
g(x)=\int_{-\frac{\epsilon}{2}}^{\frac{\epsilon}{2}}dy\,f(y)\int_{-\frac{\epsilon}{2}}^{\frac{\epsilon}{2}}dy'\, f(y')g(x;y,y'),
\eeq
where $f(y)$ is the transverse probability distribution function.

As in Secs.~\ref{sec2} and \ref{sec3}, we focus on the high-pressure behavior of $g(x;y,y')$ and $g(x)$ at sufficiently large longitudinal distances $x$. We first introduce the rescaled longitudinal distance as in Eq.~\eqref{1new}.
The choice of a single scaling value $a=\sqrt{1-\epsilon^2}$ is justified by the fact that, at high pressure, the disks tend to form a zigzag configuration, in which the longitudinal separation between neighboring disks is only slightly larger than $a$. This zigzag configuration plays the same structural role as the alternate arrangement $1$-$2$-$1$-$2$-$\cdots$ in the 1D binary case, explaining why both systems exhibit similar oscillatory behavior despite their different microscopic origins. In the limit $\bp\gg 1$, with $p$ the longitudinal pressure, the equation of state becomes $\bp\approx 2\rho/(1-\rho a)$ \cite{MS23,VBG11}, where $\rho$ is the linear density. The divergence of $\bp$ as $\rho\to 1/a$ motivates identifying the high-pressure packing fraction as $\phi=\rho a$ in Eq.~\eqref{1new}.

A careful numerical analysis of the poles of the Laplace transform of $\tg(\tx;y,y')\equiv g(x;y,y')$ (see Appendix \ref{appB} for a summary of the numerical procedure) shows that, in the high-pressure limit,
\beq
\label{28}
\zeta_n\approx  \frac{n^2\pi^2\phib^2}{4},\quad \omega_n\approx n\pi,\quad n=1,2,3,\ldots.
\eeq
Comparison with Eq.~\eqref{23} reveals that, while the oscillation frequencies follow the same pattern as in the binary mixture, the damping coefficients are half the values of the binary system with the same $\phib$. As in the factor $2$ appearing in $\bp \approx 2\phi/\phib$ \cite{MS23,VBG11}, this reduction arises from the effect of vertical fluctuations around the perfect zigzag configuration.

The ratios $\zeta_n/(n^2\pi^2\phib^2/4)$ and $\omega_n/n\pi$ for $n=1,2,\ldots,10$ are plotted in Fig.~\ref{figHD} for confined hard disks with $\epsilon=\frac{\sqrt{3}}{2}$. As in the monocomponent and binary hard-rod cases (see Figs.~\ref{fig1} and \ref{fig3}), the asymptotic expressions become less accurate as $n$ and/or $\phib$ increase, although the agreement remains very good in the high-packing regime. Interestingly, for a given void fraction $\phib$, the deviations of $\zeta_n$ and $\omega_n$ from their asymptotic high-pressure forms become progressively smaller when going from the Tonks gas to the binary mixture, and from the latter to confined hard disks. The results also confirm that the damping coefficients are systematically smaller than in the binary mixture by a factor of two, while the oscillation frequencies remain the same.

\begin{figure}
      \includegraphics[width=0.8\columnwidth]{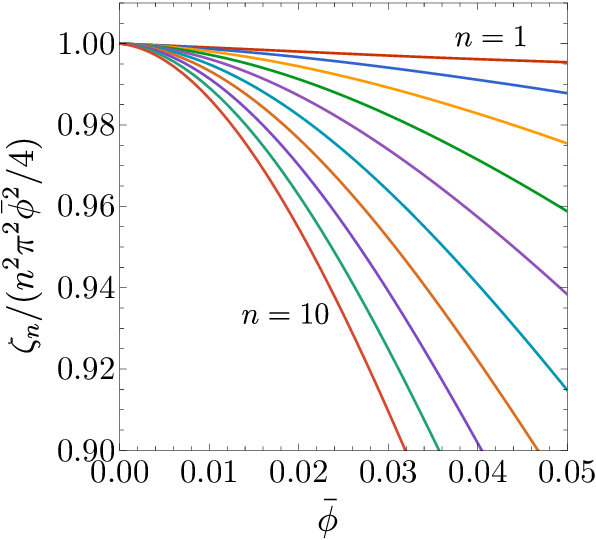}\\ \includegraphics[width=0.8\columnwidth]{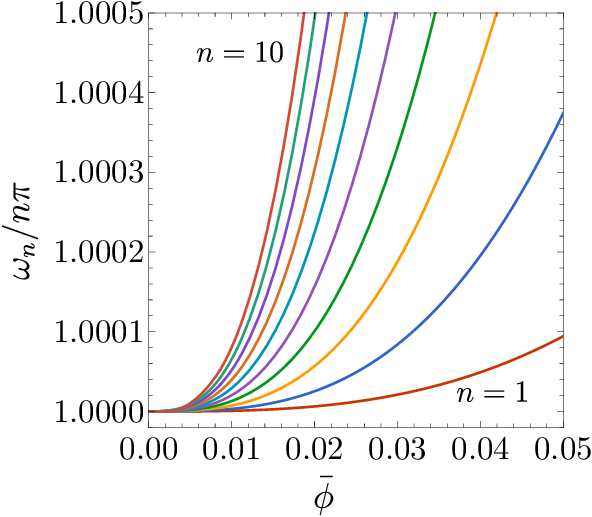}
            \caption{Confined hard disks with $\epsilon=\frac{\sqrt{3}}{2}$. (a) $\zeta_n/(n^2\pi^2\phib^2/4)$ and (b) $\omega_n/n\pi$ as functions of $\phib$ for $n=1,2,\ldots,10$ [from top to bottom in (a) and from bottom to top in (b)].
  \label{figHD}}
\end{figure}

Near close packing, the dominant RDFs are
\begin{subequations}
\beq
\tg_{++}(\tx)\equiv\tg\left(\tx;\frac{\epsilon}{2},\frac{\epsilon}{2}\right)=\tg\left(\tx;-\frac{\epsilon}{2},-\frac{\epsilon}{2}\right),
\eeq
\beq
\tg_{+-}(\tx)\equiv\tg\left(\tx;\frac{\epsilon}{2},-\frac{\epsilon}{2}\right)=\tg\left(\tx;-\frac{\epsilon}{2},\frac{\epsilon}{2}\right),
\eeq
\end{subequations}
so
\beq
\label{RDFq1D}
\tg(\tx)\approx\frac{1}{2}\left[\tg_{++}(\tx)+\tg_{+-}(\tx)\right].
\eeq
Analogously to the binary case for $\tg_{11}(\tx)$ [Eq.~\eqref{25a}], the residues of the Laplace transform of $\tg_{++}(\tx)$ alternate in sign.

In summary, for $\phib\ll 1$ we have
\begin{subequations}
\label{29}
\beq
\tg_{++}(\tx)\approx\theta_3\left(\frac{\pi(\tx-1)}{2},e^{-\pi^2\phib^2\tx/4}\right),
\eeq
\beq
\tg_{+-}(\tx)\approx\theta_3\left(\frac{\pi\tx}{2},e^{-\pi^2\phib^2\tx/4}\right),
\eeq
\beq
\tg(\tx)
\approx\theta_3\left(\pi\tx,e^{-\pi^2\phib^2\tx}\right).
\eeq
\end{subequations}
The corresponding envelopes are
\begin{subequations}
\beq
\tge_{++}(\tx)
=\tge_{+-}(\tx)=\theta_3\left(0,e^{-\pi^2\phib^2\tx/4}\right),
\eeq
\beq
\tge(\tx)=\theta_3\left(0,e^{-\pi^2\phib^2\tx}\right).
\eeq
\end{subequations}
As a consequence,
\begin{subequations}
\beq
\label{alg++}
\tge_{++}(\tx) \simeq
\begin{cases}
\displaystyle{\frac{2}{\sqrt{\pi\phib^2\tx}}},& \displaystyle{\tx\lesssim \frac{0.62}{\phib^2}}\\
\displaystyle{1+2e^{-\pi^2\phib^2\tx/4}},&\displaystyle{\tx\gtrsim \frac{0.62}{\phib^2}},
\end{cases}
\eeq
\beq
\label{algHD}
\tge(\tx) \simeq
\begin{cases}
 \displaystyle{\frac{1}{\sqrt{\pi\phib^2\tx}}},& \displaystyle{\tx\lesssim \frac{0.15}{\phib^2}}\\
 \displaystyle{1+2e^{-\pi^2\phib^2\tx}},&\displaystyle{\tx\gtrsim \frac{0.15}{\phib^2}}.
\end{cases}
\eeq
\end{subequations}

\begin{figure}
      \includegraphics[width=0.8\columnwidth]{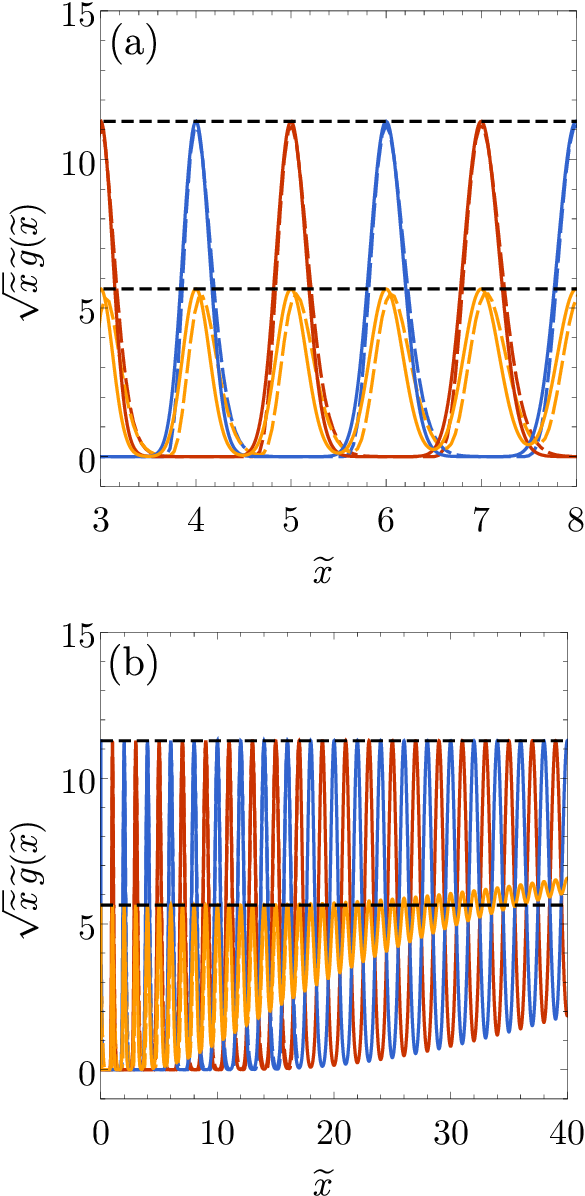}
            \caption{Confined hard disks with $\epsilon=\frac{\sqrt{3}}{2}$.  Plot of $\sqrt{\tx}\tg_{++}(\tx)$ (maxima at $\tx\approx 1,3,5,\ldots$), $\sqrt{\tx}\tg_{+-}(\tx)$ (maxima at $\tx\approx 0,2,4,\ldots$), and $\sqrt{\tx}\tg(\tx)$ (maxima at $\tx\approx 0,1,2,\ldots$) for $\phi=0.9$. Solid lines correspond to the exact functions, and dashed lines to our approximations [Eqs.~\eqref{29}]. The horizontal dashed lines indicate $2/\sqrt{\pi\phib^2}\simeq 11.28$ and $1/\sqrt{\pi\phib^2}\simeq 5.64$. (a) $3\leq\tx\leq 8$. (b) $0\leq\tx\leq 40$.
  \label{fig5}}
\end{figure}

Figure \ref{fig5} demonstrates the good agreement between the exact distributions $\tg_{++}(\tx)$, $\tg_{+-}(\tx)$, and $\tg(\tx)$ and our theoretical predictions, Eqs.~\eqref{29}. Comparison with Fig.~\ref{fig4} shows that the regime of intermediate algebraic decay extends over a wider range for confined disks than for the binary mixture with the same $\phi$ and $a$. We also note that the agreement between the total RDF $\tg(\tx)$ and its approximation is slightly less accurate in the q1D confined disk system than in the 1D binary mixture, which is expected since Eq.~\eqref{RDFq1D} introduces an additional approximation in the q1D case.

The intermediate algebraic decay $g^{\text{env}}(x) \sim 1/\sqrt{x}$ for confined disks was previously estimated in Ref.~\cite{HC21}, improving upon the earlier decay law $g^{\text{env}}(x) \sim x^{-2/3}$ proposed from computer simulations in Ref.~\cite{HBPT20}.
Hu and  Charbonneau (HC) justified their result by starting from the 1D algebraic decay [Eq.~\eqref{alg}], which in terms of pressure reads $g^{\text{env}}(x)\simeq (1+\bp)^{3/2}/\sqrt{2\pi \bp (x-1)}$.
They then made the substitutions $\bp\to\bp a$ and $x\to x/a$, and introduced a correction factor $c$, obtaining
\beq
\label{HC}
g^{\text{env}}_{\text{HC}}(x)\simeq c \frac{(1+\bp a)^{3/2}}{\sqrt{2\pi \bp a (x/a-1)}}.
\eeq
By fitting data obtained via the planting method, they estimated $c = 0.7$. Using $\bp a \simeq 2\phi/\phib$ and the scaled distance in Eq.~\eqref{1new}, Eq.~\eqref{HC} can be rewritten as
\beq
\label{HCbis}
\tge_{\text{HC}}(\tx)\simeq c \sqrt{2}\frac{(1-\phib/2)^{3/2}}{\sqrt{\pi \phib^2\tx}}.
\eeq
Setting $1-\phib/2\simeq 1$ and choosing $c = 1/\sqrt{2} \simeq 0.707$, Eq.~\eqref{HCbis} exactly reproduces our result in Eq.~\eqref{algHD}. Thus, our approach removes the need for empirical fitting entirely.

\section{Conclusions}
\label{sec5}

In this paper we have obtained exact asymptotic expressions for the RDFs of 1D and q1D hard-core systems near close packing and expressed them in a compact form determined by the pole structure of their Laplace transforms. The resulting representation, given in terms of Jacobi theta functions, not only provides a concise description in terms of a single special function, but also offers direct physical insight by explicitly linking spatial correlations to the Laplace-transform poles: each pole $s_n=-\zeta_n\pm \imath\omega_n$ defines a decay mode with a characteristic damping rate and oscillation frequency.

In contrast to earlier approaches that focused primarily on the decay of the envelope of $g(x)$, our analysis yields closed-form expressions for the full oscillatory RDFs and reveals mode-filtering effects that are not captured by envelope-based descriptions. The crossover behavior between algebraic and exponential decay in the correlations of a Tonks gas was recently obtained by Bouzar and Messina using a different approach, in which $\tg(\tx)$ is decomposed into Gaussian peaks centered at $\tx=n$. By comparison, the present formulation [Eq.~(\ref{Myg})] represents $\tg(\tx)$ as a superposition of extended, damped oscillatory modes. While both approaches recover the same envelope decay, the Laplace-pole representation additionally resolves the oscillatory structure and its associated decay mechanisms.

Although the Tonks gas provides a useful reference case, the scope of the present results extends well beyond known 1D behavior. The pole-analysis framework applies naturally to more complex systems and leads to analytical results for symmetric nonadditive binary mixtures and for q1D confined hard disks, where previous studies have been largely limited to numerical analyses of envelope decay. In these systems, our method yields explicit expressions not only for the total RDF (including both decay and oscillatory components) but also for partial correlation functions. These partial correlations display distinct oscillatory patterns and different algebraic and exponential decay regimes that have remained unexplored.

A synoptic comparison of the three systems---Tonks gas, binary mixture, and confined hard disks---is provided in Table~\ref{tab1}, which places them on equal footing and highlights their common pole structure together with the quantitative differences in decay rates, crossover scales, and mode selection induced by composition or confinement. The pole structure obtained in the high-packing-fraction limit shows that, for all systems considered, the decay rates scale as $\zeta_n\propto n^2\phib^2$ for $\phib\ll 1$, reflecting the increasingly ordered, quasi-crystalline arrangements near close packing. In the case of confined hard disks, the present pole-based derivation fixes all numerical prefactors exactly, removing the need for heuristic substitutions or empirical fitting parameters introduced in earlier treatments.

The common feature underlying all these cases is the emergence of an alternating or quasi-periodic arrangement near close packing, which effectively reduces the problem to a single-file ordering mechanism. This same structural mechanism also appears in other confined systems, such as freely rotating hard dumbbells under strong confinement \cite{MGVS26}. It explains why the same algebraic-to-exponential crossover arises in otherwise different models, while the remaining quantitative differences reflect the specific geometrical constraints, effective contact distances, and close-packing equations of state of each system.

More generally, the analysis demonstrates that the nature of correlation decay near close packing is governed by the spectrum of Laplace poles, providing a systematic alternative to real-space decompositions and asymptotic matching arguments. The same framework can be extended to time-dependent correlation and response functions in single-file and quasi-single-file geometries, where the poles control both spatial and temporal decay. The results presented here thus provide a coherent foundation for future studies of correlations and dynamics in strongly confined 1D and q1D hard-core fluids.

\appendix
\section{Laplace transforms for symmetric nonadditive binary mixtures}
\label{appA}
The Laplace transforms of $g_{ij}(x)$ are \cite{S16}
\begin{subequations}
\beq
G_{11}(s)=G_{22}(s)=\frac{2}{\rho}\frac{1-Q_{11}(s)-D(s)}{D(s)},
\eeq
\beq
G_{12}(s)=\frac{2}{\rho}\frac{Q_{12}(s)}{D(s)},
\eeq
\end{subequations}
where
\begin{subequations}
\beq
Q_{11}(s)=Q_{22}(s)=\frac{\bp}{s+\bp}\frac{e^{-s}}{1+e^{(1-a)\bp}},
\eeq
\beq
Q_{12}(s)=\frac{\bp}{s+\bp}\frac{e^{-a s}}{1+e^{-(1-a)\bp}},
\eeq
\bal
D(s)=&\left[1-\frac{\bp}{s+\bp}\frac{e^{-s}}{1+e^{(1-a)\bp}}\right]^2\nn
&-\left[\frac{\bp}{s+\bp}\frac{e^{-a s}}{1+e^{-(1-a)\bp}}\right]^2.
\eal
\end{subequations}

In the high-pressure regime ($\bp\gg 1$),
\begin{subequations}
\beq
Q_{11}(s)\approx 0,\quad Q_{12}(s)\approx\left(1+\frac{a \phib s}{\phi}\right)^{-1} e^{-as},
\eeq
\beq
D(s)\approx\frac{1+\frac{a \phib }{\phi}s-e^{- as}}{1+\frac{a \phib }{\phi}s}\frac{1+\frac{a \phib }{\phi}s+e^{- as}}{1+\frac{a \phib }{\phi}s}.
\eeq
\end{subequations}

\section{Laplace transforms for confined hard disks}
\label{appB}

In this Appendix we summarize the main expressions yielding the Laplace transform $G(s;y,y')$ of the partial RDF $g(x;y,y')$. The reader is referred to Refs.~\cite{MS23b,MS24,MGVS26} for full derivational details.

The transverse probability distribution function $f(y)$ is obtained from the eigenvalue problem
\beq
\int_{-\frac{\epsilon}{2}}^{\frac{\epsilon}{2}} dy'\,\Omega(\bp;y,y')\sqrt{f(y')}=\lambda_0\sqrt{f(y)},
\eeq
where $\lambda_0$ is the largest eigenvalue and
\beq
\Omega(s;y,y')=\frac{e^{-s\sigma(y,y')}}{s}.
\eeq

Next, we introduce the auxiliary Laplace function
\beq
\label{IE}
\Gamma(s;y,y')\equiv \rho \sqrt{f(y)f(y')}G(s;y,y'),
\eeq
which satisfies the integral equation
\bal
\label{32}
\lambda_0\Gamma(s;y,y')=
&\int_{-\frac{\epsilon}{2}}^{\frac{\epsilon}{2}} dy''\, \Gamma(s;y,y'')\Omega(s+\bp;y'',y')\nn
&+\Omega(s+\bp;y,y').
\eal
In operator form, this can be written as
\beq
\label{34}
\mathbf{\Gamma}(s)=\left[\lambda_0\mathbf{I}-\mathbf{\Omega}(s+\bp)\right]^{-1}\cdot \mathbf{\Omega}(s+\bp),
\eeq
where $\mathbf{\Gamma}(s)$ and $\mathbf{\Omega}(s)$ are operators whose kernels are $\Gamma(s;y,y')$ and $\Omega(s;y,y')$, respectively, and $\mathbf{I}$ is the identity operator.

The Laplace transform of $\tg(\tx;y,y')$ is related to $G(s;y,y')$ through $\tG(s;y,y')=(\phi/a)e^{\phi s}G(\phi s/a;y,y')$. In practice, the poles of $\tG(s;y,y')$ are obtained in two steps. First, the transverse coordinate $y$ is discretized into $M$ values, so that Eq.~\eqref{34} becomes a finite matrix equation. The roots of the determinant of the discrete $M\times M$ matrix $\lambda_0\mathbf{I}-\mathbf{\Omega}(\phi s/a+\bp)$ are computed numerically for several, large enough values of $M$. Second, these roots are extrapolated to the continuum limit $M\to\infty$.

\section*{Acknowledgments}
The authors acknowledge financial support from Grant No.~PID2024-156352NB-I00 funded by MCIU/AEI/10.13039/501100011033 and by ERDF/EU, and from Grant No.~GR24022 funded by the Junta de Extremadura (Spain), A.M.M. is grateful to the Spanish Ministerio de Ciencia e Innovaci\'on for Fellowship No. PRE2021-097702.

\section*{Data availability} Data supporting figures and numerical results are available from the corresponding author upon reasonable request.

\end{document}